# Exploiting Structure in Cooperative Bayesian Games

**Frans A. Oliehoek**
Dept. of Knowledge Engineering
Maastricht University
Maastricht, The Netherlands

**Shimon Whiteson**
Informatics Institute
University of Amsterdam
Amsterdam, The Netherlands

**Matthijs T.J. Spaan**
Faculty EEMCS
Delft University of Technology
Delft, The Netherlands

## Abstract

Cooperative *Bayesian games* (BGs) can model decision-making problems for teams of agents under imperfect information, but require space and computation time that is exponential in the number of agents. While *agent independence* has been used to mitigate these problems in perfect information settings, we propose a novel approach for BGs based on the observation that BGs additionally possess a different types of structure, which we call *type independence*. We propose a factor graph representation that captures both forms of independence and present a theoretical analysis showing that *non-serial dynamic programming* cannot effectively exploit type independence, while MAX-SUM can. Experimental results demonstrate that our approach can tackle cooperative Bayesian games of unprecedented size.

## 1 Introduction

Cooperative multiagent systems are an essential tool, not only for tackling inherently distributed problems, but also for decomposing problems too complex to be tackled by a single agent (Huhns, 1987; Sycara, 1998; Panait and Luke, 2005; Vlassis, 2007; Buşoniu et al., 2008). However, to make such systems effective, the constituent agents must be able to coordinate their action selection in order to achieve a common goal.

This problem is particularly vexing in the presence of *imperfect information* (Harsanyi, 1967–1968). Even a single-agent system may have incomplete knowledge of the state of its environment (Kaelbling et al., 1998), e.g., due to noisy sensors. However, the presence of multiple agents often greatly exacerbates this problem, as each agent typically has access to only a fraction of the whole system's sensors. In principle, agents could synchronize their beliefs and coordinate their actions by communicating. However, communication is often unreliable and limited by bandwidth constraints, leaving each agent uncertain how others will act.

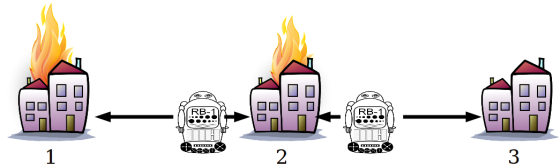

Figure 1: Illustration of multiagent decision making with imperfect information. Both agents know house 2 is on fire. However, each agent receives only a noisy observation of the single neighboring house it can observe in the distance.

Consider the example depicted in Fig. 1. Both firefighting agents know there is fire at house $H_2$ but each receives only a noisy observation about one of the other houses. Consequently, effective decision making is difficult. If agent 1 (left) observes flames in house $H_1$, it may be tempted to fight fire there rather than at house $H_2$. However, the efficacy of doing so depends on whether agent 2 (right) will fight fire in house $H_2$, which in turn depends on whether agent 2 observes flames in house $H_3$, a fact unknown to agent 1.

Such problems can be modeled as cooperative *Bayesian games (BGs)*, which extend traditional *strategic games* to model imperfect information. BGs can model a great variety of problems and are of great importance for solving sequential decision problems (Emery-Montemerlo et al., 2004; Oliehoek et al., 2008a; Kumar and Zilberstein, 2010; Spaan et al., 2011). In a BG, a *type* for each agent, specifying the private information it holds, is drawn from a distribution. Then, the agents simultaneously select actions conditioned on their individual type, and subsequently receive a team payoff based on the realized types and actions.

Unfortunately, solving cooperative BGs is NP-hard (Tsitsiklis and Athans, 1985). Just representing them requires space exponential in the number of agents and solving them exactly requires time exponential in the number of agents and types. However, as in cooperative strategic games (Guestrin et al., 2002; Kok and Vlassis, 2006), many problems exhibit *agent independence*, i.e., the global payoff function can be decomposed as the sum of *local payoff functions*, each of which depends on the actions of only a few agents. The resulting *cooperative graphical Bayesian games (CG-BGs)* can be represented more compactly and solved more effectively using inference methods for graphical models such as *non-serial dynamic programming* (NDP) (Bertele and Brioschi, 1972) and the popular MAX-SUM algorithm (Pearl, 1988; Kok and Vlassis, 2005, 2006; Farinelli et al., 2008; Rogers et al., 2011).

This paper is motivated by the observation that BGs also exhibit another form of structure that we call *type independence*, in which the global payoff function can be decomposed as the sum of *contributions*, each of which depends on only one joint type. Such structure has been exploited when cooperative BGs are solved using heuristic search methods (Kumar and Zilberstein, 2010; Oliehoek et al., 2010), though the role of type independence has not been made explicit. However, in the worst case, heuristic search methods are no better than brute-force search and thus can perform much worse than inference-based methods, even when exploiting structure with *and/or search trees* (Marinescu and Dechter, 2009).[1] In addition, such methods do not exploit agent independence.

To address these shortcomings, this paper investigates the exploitation of type independence using inference methods. Specifically, we propose a novel *factor graph* representation that neatly captures *both* agent and type independence and thus enables the exploitation of both forms of structure. Our analysis of the computational complexity of both NDP and MAX-SUM applied to this factor graph reveals that the former cannot effectively exploit type independence; we prove that its computational complexity remains exponential in the number of types (although it still outperforms one of the search methods empirically). On the other hand, we prove that each iteration of MAX-SUM is tractable for problems with small local neighborhoods. Bounding the number of iterations thus yields a polynomial-time approximate method for such cooperative BGs.

In addition, we present extensive experimental results that demonstrate that MAX-SUM finds near-optimal solutions and that the simultaneous exploitation of both agent and type independence leads to dramatically better scalability than several state-of-the-art alternatives in the number of agents, actions, and types. This improved scalability enables the solution of cooperative BGs of unprecedented size.

## 2 Background

We begin with some background on BGs and CGBGs.

**Bayesian Games** A Bayesian game, also called a *strategic game of imperfect information*, is an augmented strategic game in which the players hold private information (Harsanyi, 1967–1968; Osborne and Rubinstein, 1994). The private information of agent $i$ defines its *type* $\theta_i \in \Theta_i$. The agents' payoffs depend not only on their actions, but also on their types.

**Definition 1.** A *Bayesian game (BG)* is a tuple $\langle \mathcal{D}, \mathcal{A}, \boldsymbol{\Theta}, \Pr(\boldsymbol{\Theta}), \langle u_1, ...u_n \rangle \rangle$, where $\mathcal{D}$ is the set of $n$ agents, $\mathcal{A} = \mathcal{A}_1 \times \cdots \times \mathcal{A}_n$ is the set of joint actions $\mathbf{a} = \langle a_1, \ldots, a_n \rangle$, $\boldsymbol{\Theta} = \Theta_1 \times \cdots \times \Theta_n$ is the set of joint types $\boldsymbol{\theta} = \langle \theta_1, \ldots, \theta_n \rangle$, $\Pr(\boldsymbol{\Theta})$ is the distribution over joint types, and $u_i : \boldsymbol{\Theta} \times \mathcal{A} \to \mathbb{R}$ is the payoff function of agent $i$.

Since agents in a BG can condition their actions on their types, they select policies (as opposed to actions as in strategic games). A joint policy $\boldsymbol{\beta} = \langle \beta_1, ..., \beta_n \rangle$, consists of individual policies $\beta_i$ for each agent $i$. Deterministic (pure) individual policies map each type $\theta_i$ to an action. Since we focus on *cooperative* Bayesian games, in which all agents share a single team payoff function $u = u_i$ for all $i$, only deterministic policies need be considered.

In a BG, the traditional Nash equilibrium is replaced by a *Bayesian Nash equilibrium (BNE)*. A profile of policies $\boldsymbol{\beta} = \langle \beta_1, ..., \beta_n \rangle$ is a BNE when no agent $i$ has an incentive to switch its policy $\beta_i$, given the policies of the other agents $\boldsymbol{\beta}_{\neq i}$. In cooperative BGs, the definition of a solution is simpler. Let the *value* of a joint policy be its expected payoff:

$$V(\boldsymbol{\beta}) = \sum_{\boldsymbol{\theta} \in \boldsymbol{\Theta}} \Pr(\boldsymbol{\theta}) u(\boldsymbol{\theta}, \boldsymbol{\beta}(\boldsymbol{\theta})), \qquad (1)$$

where $\boldsymbol{\beta}(\boldsymbol{\theta}) = \langle \beta_1(\theta_1), ..., \beta_n(\theta_n) \rangle$ is the joint action specified by $\boldsymbol{\beta}$ for joint type $\boldsymbol{\theta}$. A solution of a cooperative BG is $\boldsymbol{\beta}^* = \arg\max_{\boldsymbol{\beta}} V(\boldsymbol{\beta})$, which is a Pareto-optimal Bayesian Nash equilibrium.

**Cooperative Graphical Bayesian Games.** A key difficulty in BGs is that the size of their payoff matrices is exponential in the number of agents. However, many BGs contain *agent independence*, i.e., not all agents directly influence each other, which allows for compact representations and efficient solutions.

---

[1] NDP, for example, is never slower than brute-force search, i.e., conditioning in graphical models (Koller and Friedman, 2009). When the *induced width* is small, NDP can be much faster (Dechter, 1999).

Agent independence can be formalized in *graphical (strategic) games* (Kearns et al., 2001; Kearns, 2007) and *graphical Bayesian games* (Soni et al., 2007), in which each agent has an *individual* payoff function that depends on only a subset of agents. Unfortunately, such models are not useful in cooperative games since all agents share a single payoff function in which all agents participate (otherwise they would be irrelevant). Instead, we employ *cooperative graphical BGs (CGBGs)* (Oliehoek et al., 2008b), which use *coordination (hyper-)graphs* (Guestrin et al., 2002; Kok and Vlassis, 2006) to express conditional independence between agents. In CGBGs, the single team payoff function decomposes as the sum of several *local* payoff functions.

**Definition 2.** A *cooperative graphical Bayesian game (CGBG)* is a tuple $\langle \mathcal{D}, \mathcal{A}, \Theta, \Pr(\Theta), \mathcal{U} \rangle$ where $\mathcal{D}, \mathcal{A}, \Theta,$ and $\Pr(\Theta)$ are as before and $\mathcal{U} = \{u^1, \ldots, u^\rho\}$ is the set of *local payoff functions*, each of which involves only a subset of agents. Each $u^e$ has *scope* $\mathbb{A}(u^e)$ specifying the subset of agents on which it depends.

When scopes are restricted, CGBGs can be represented much more compactly than regular cooperative BGs. In fact, if the largest scope has size $k$, then the representation size is exponential in $k$, but only linear in $n$. However, the joint type distribution must also be compact. A typical assumption is that the type probabilities factor as the product of individual type probabilities (Soni et al., 2007; Jiang and Leyton-Brown, 2010). More generally, it is common to represent $\Pr(\Theta)$ compactly (e.g., as in (Koller and Milch, 2003; Jiang and Leyton-Brown, 2010)) by means of Bayesian networks (Pearl, 1988; Bishop, 2006) or other graphical models.[2]

Agents in a CGBG aim to maximize the expected sum of local payoffs:

$$\boldsymbol{\beta}^* = \arg\max_{\boldsymbol{\beta}} \sum_{e \in \mathcal{E}} \sum_{\boldsymbol{\theta}_e \in \Theta_e} \Pr(\boldsymbol{\theta}_e) u^e(\boldsymbol{\theta}_e, \boldsymbol{\beta}_e(\boldsymbol{\theta}_e)) \quad (2)$$

where $\boldsymbol{\theta}_e = \langle \theta_i \rangle_{i \in \mathbb{A}(u^e)}$ is the local joint type, $\boldsymbol{\beta}_e = \langle \beta_i \rangle_{i \in \mathbb{A}(u^e)}$ is the local joint policy, and $\boldsymbol{\beta}_e(\boldsymbol{\theta}_e)$ is the local joint action under $\boldsymbol{\beta}$ given $\boldsymbol{\theta}_e$.[3] The local value for the $e$-th payoff component is:

$$V^e(\boldsymbol{\beta}_e) = \sum_{\boldsymbol{\theta}_e \in \Theta_e} \Pr(\boldsymbol{\theta}_e) u^e(\boldsymbol{\theta}_e, \boldsymbol{\beta}_e(\boldsymbol{\theta}_e)). \quad (3)$$

---

[2]While not all joint type distributions admit a compact representation, this is nonetheless a minimal assumption: there is no hope of solving other BGs efficiently, as they cannot even be represented efficiently. Fortunately, the class of real-world problems that admit a compact representation is likely to be quite large.

[3]We abuse notation such that $e$ is an index into the set of local payoff functions and an element of the set of scopes.

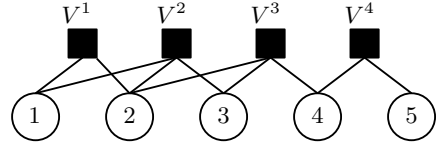

Figure 2: An AI factor graph with five agents. Circular nodes are agents and square nodes are local payoff functions, with edges indicating in which local payoff function each agent participates.

Each local value component can be interpreted as a factor in a *factor graph* (Kschischang et al., 2001; Loeliger, 2004), which generalizes a coordination graph to allow local payoff functions that depend on more than two agents. The resulting bipartite graph has one set of nodes for all the local value functions (the factors) and another set for all the agents (the variables, whose values correspond to policies), as shown in Fig. 2. A local value function $V^e$ is connected to an agent $i$ if and only if $i \in \mathbb{A}(u^e)$. We refer to this as an *agent independence (AI) factor graph*.

A naive solution approach is to simply enumerate all joint policies. However, the required time is $O(|\mathcal{A}_*|^{|\Theta_*|^n})$, where $|\mathcal{A}_*|$ and $|\Theta_*|$ are the size of the largest individual action and type set. Thus, it scales exponentially with the number of agents and types. Fortunately, the agent independence expressed in the factor graph of Fig. 2 can be exploited. For instance, Oliehoek et al. (2008b) apply NDP to this graph. Another option is to use MAX-SUM message passing.

## 3 Representing Type Independence

Applying NDP or MAX-SUM to an AI factor graph can reduce the exponential complexity with respect to the number of agents. However, it cannot reduce the exponential complexity with respect to the number of types because this complexity manifests itself in the number of values that the variables can take on.

The key observation motivating this work is that CGBGs also possess a second form of independence, which we call *type independence*, that enables solution approaches to avoid the exponential dependence on the number of types. Unlike agent independence, which only some BGs possess, type independence is an inherent property of all BGs because it is a direct consequence of the fact that the value of a joint policy is an expectation over joint types.

Type independence is expressed in terms of a *contribution* for each joint type:

$$C_{\boldsymbol{\theta}}(\mathbf{a}) \equiv \Pr(\boldsymbol{\theta}) u(\boldsymbol{\theta}, \mathbf{a}). \quad (4)$$

The value of a joint policy $\boldsymbol{\beta}$ can now be interpreted

as a sum of contributions, one for each joint type:

$$V(\boldsymbol{\beta}) = \sum_{\boldsymbol{\theta} \in \boldsymbol{\Theta}} C_{\boldsymbol{\theta}}(\boldsymbol{\beta}(\boldsymbol{\theta})). \qquad (5)$$

Type independence results from the additive structure of a joint policy's value shown in (5). The key insight is that each contribution term depends only on the joint action selected when the corresponding joint type $\boldsymbol{\theta}$ occurs. Since that action is selected according to the joint policy $\boldsymbol{\beta}$, the contribution depends only on $\boldsymbol{\beta}(\boldsymbol{\theta})$, the part of the joint policy that specifies the joint action for $\boldsymbol{\theta}$.

Viewed from another perspective, type independence occurs because, in any game, only one individual type is realized for each agent, and that type affects only some contributions. In other words, the individual action $\beta_i(\theta_i)$ selected for type $\theta_i$ of agent $i$ affects only those contributions whose joint types involve $\theta_i$. For instance, in the firefighting problem illustrated in Fig. 1, one possible joint type is $\boldsymbol{\theta} = \langle N, N \rangle$ (neither agent observes flames). Clearly, the action $\beta_1(F)$ that agent 1 selects when it has type $F$ (it observes flames), has no effect on the contribution of this joint type.

An optimal joint policy can also be described in terms of contributions, by simply maximizing the value as defined in (5):

$$\boldsymbol{\beta}^* = \arg\max_{\boldsymbol{\beta}} V(\boldsymbol{\beta}) = \arg\max_{\boldsymbol{\beta}} \sum_{\boldsymbol{\theta} \in \boldsymbol{\Theta}} C_{\boldsymbol{\theta}}(\boldsymbol{\beta}(\boldsymbol{\theta})). \qquad (6)$$

Thus, the solution to a BG maximizes the sum of contributions, each of which has restricted scope. This is an important observation, since maximization of an additively factored function with components of restricted scope is exactly the operation that algorithms such as NDP and MAX-SUM make more efficient.

To apply these methods, we represent type independence in a factor graph, by creating factors for all the contributions (corresponding to joint types) and variables for all the individual types of all the agents. Fig. 3 shows such a factor graph for the firefighting problem. Unlike the representation that results from reducing a BG to a strategic game played by *agent-type* combinations (Osborne and Rubinstein, 1994), this factor graph does not 'flatten' the utility function. On the contrary, it explicitly represents the contributions of each joint type, thereby capturing type independence.

However, while the factor graph in Fig. 3 represents type independence, it does not represent agent independence because the global payoff represented by the sum of contributions is not decomposed into local payoff functions. Conversely, the factor graph shown in Fig. 2 represents agent independence but not type in-dependence because the value of each local payoff function is not decomposed into contributions. To solve CGBGs efficiently, we need a factor graph formulation that captures both agent and type independence.

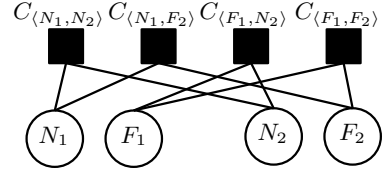

Figure 3: A factor graph for the firefighting problem that captures type independence, e.g., the action that agent 1 selects when it has type $N$ affects only the contribution factors $C_{\langle N_1, N_2 \rangle}$ and $C_{\langle N_1, F_2 \rangle}$ in which it has that type.

To this end, we define a *local contribution* as follows:

$$C^e_{\boldsymbol{\theta}_e}(\mathbf{a}_e) \equiv \Pr(\boldsymbol{\theta}_e) u^e(\boldsymbol{\theta}_e, \mathbf{a}_e). \qquad (7)$$

Using this notation, the solution of the CGBG is

$$\boldsymbol{\beta}^* = \arg\max_{\boldsymbol{\beta}} \sum_{e \in \mathcal{E}} \sum_{\boldsymbol{\theta}_e \in \boldsymbol{\Theta}_e} C^e_{\boldsymbol{\theta}_e}(\boldsymbol{\beta}_e(\boldsymbol{\theta}_e)). \qquad (8)$$

Thus, the solution corresponds to the maximum of an additively decomposed function containing a contribution for each local joint type $\boldsymbol{\theta}_e$. This can be expressed in a factor graph in which an individual type $\theta_i$ of an agent $i$ is connected to a contribution $C^e_{\boldsymbol{\theta}_e}$ if and only if $i$ participates in $u^e$ and $\boldsymbol{\theta}_e$ specifies $\theta_i$ for agent $i$, as illustrated in Fig. 4. We refer to this graph as the *agent and type independence (ATI) factor graph*. Contributions are separated, not only by the joint type to which they apply, but also by the local payoff function to which they contribute. Consequently, both agent and type independence are neatly expressed.

## 4 Solving ATI Factor Graphs

ATI factor graphs can be solved using existing methods such as NDP and MAX-SUM. In this section, we analyze the computational complexity of doing so.

### 4.1 Non-Serial Dynamic Programming

NDP can compute the maximum configuration of a factor graph. In the *forward pass*, variables are eliminated one by one according to some prespecified order. Eliminating the $k$th variable $v$ involves collecting all the factors in which it participates and replacing them with a new factor $f^k$ that represents the sum of the removed factors, given that $v$ selects a best response. Once all variables are eliminated, the *backward pass* iterates through the variables in reverse order of elimination. Each variable selects a best response to the variables already visited, eventually yielding an optimal joint policy.

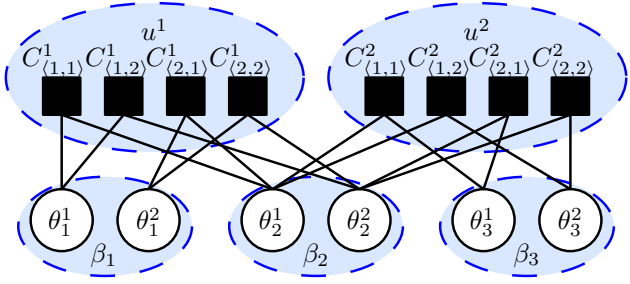

Figure 4: ATI factor graph for a CGBG with three agents, two types per agent, and two local payoff functions. Agents 1 and 2 participate in payoff function $u^1$ (corresponding to the first four contributions), while agents 2 and 3 participate in $u^2$ (corresponding to the last four contributions). The factor graph expresses both agent independence (e.g., agent 1 does not participate in $u^2$) and type independence (e.g, the action agent 1 selects when it receives observation $\theta_1^1$ affects only the first 2 contributions).

The maximum number of agents participating in a factor encountered during NDP is known as the induced width $w$ of the ordering. The induced width depends on the order in which the variables are eliminated and determining the optimal order is NP-complete (Arnborg et al., 1987). The following result is well-known (see for instance (Dechter, 1999)):

**Theorem 1.** NDP *requires exponential time and space in the induced width $w$.*

Though NDP is still exponential, for sparse problems the induced width is much smaller than the total number of variables $V$, i.e., $w \ll V$, leading to an exponential speedup over naively enumerating the joint variables. However, NDP scales poorly in practice on densely connected graphs (Kok and Vlassis, 2006). In addition, because of the particular shape that type independence induces on the ATI factor graph, we can establish the following:

**Theorem 2.** *The induced width of an ATI factor graph is lower bounded by the number of individual types and the size of the largest payoff function:*

$$w \geq (k-1) \cdot |\Theta_*|,$$

*where $k$ is the largest local scope $k = \max_{e \in \mathcal{E}} |\mathbb{A}(u^e)|$, and $\Theta_*$ denotes the smallest individual type set.*

*Proof.* Assume $u^e$ is a payoff function of maximal size $k$. At some point, NDP will eliminate the first variable connected to one of its contributions, i.e., a type of an agent $i \in \mathbb{A}(u^e)$. Since such a variable is connected to contributions for each profile $\boldsymbol{\theta}_{e \setminus i}$ of types of the other agents participating in $u^e$, the newly introduced factor will connect all the types of these $(k-1)$ other agents. The minimum number of types of an agent is $|\Theta_*|$ and thus the lower bound holds. $\square$

Theorems 1 and 2 lead directly to the following observation:

**Corollary 1.** *The computational complexity of NDP applied to an ATI factor graph is exponential in the number of individual types.*

Therefore, even given the ATI factor graph formulation, it seems unlikely that NDP can effectively exploit type independence. In particular, we hypothesize that NDP will not perform significantly better when applied to the ATI factor graph instead of the AI factor graph. In fact, it is easy to construct an example where it performs worse.

### 4.2 Max-Sum

To more effectively exploit type independence, we can solve the ATI factor graph using the MAX-SUM message passing algorithm. MAX-SUM is an appealing choice for several reasons. First, it performs well in practice on structured problems (Murphy et al., 1999; Kschischang et al., 2001; Kok and Vlassis, 2006; Farinelli et al., 2008; Kuyer et al., 2008). Second, unlike NDP, it is an *anytime* algorithm that can provide results after each iteration, not only at the end (Kok and Vlassis, 2006). Third, as we show below, its computational complexity is exponential only in the size of the largest local payoff function's scope, which is fixed for many classes of CGBGs.

MAX-SUM iteratively sends messages between the factors and variables. These messages encode how much payoff the sender expects to be able to contribute to the total payoff. In particular, a message sent from a type $i$ to a contribution $j$ encodes, for each possible action, the payoff it expects to contribute. This is computed as the sum of the incoming messages from other contributions. Similarly, a message sent from a contribution to a type $i$ encodes the payoff it can contribute conditioned on each available action to the agent with type $i$.[4]

MAX-SUM iteratively passes these messages over the edges of the factor graph. Within each iteration, the messages are sent either in parallel or sequentially with a fixed or random ordering. When run on an acyclic factor graph (i.e., a tree), it is guaranteed to converge to an optimal fixed point (Pearl, 1988; Wainwright et al., 2004). In cyclic factor graphs, there are no guarantees that MAX-SUM will converge.[5] However, experimental results have demonstrated that it works well in practice even when cycles are present (Kschischang et al., 2001; Kok and Vlassis, 2006; Kuyer et al., 2008). This requires normalizing the messages to prevent them from growing ever larger, e.g., by taking a weighted sum of the new and old messages (damping).

---
[4] For a detailed description of how the messages are computed, see Oliehoek (2010) or Rogers et al. (2011).

Here we show that the computational complexity of one iteration of MAX-SUM on a CGBG is exponential only in the size of the largest local payoff function's scope. In general, it is not possible to bound the number of iterations, since MAX-SUM is not guaranteed to converge. However, MAX-SUM converges quickly in practice and, since it is an anytime algorithm, the number of iterations can be fixed in advance.

**Theorem 3.** *One iteration of* MAX-SUM *run on the factor graph constructed for a CGBG has cost*

$$O\left(|\mathcal{A}_*|^k \cdot k^2 \cdot \rho \rho_* |\Theta_*|^{2k-1}\right), \quad (9)$$

*where $\rho_*$ is the maximum number of edges in which an agent participates.*

*Proof.* It follows directly from the cost of a message and the number of messages sent by factors and variables that the complexity of one iteration of MAX-SUM is

$$O\left(m^k \cdot k^2 \cdot l \cdot F\right), \quad (10)$$

where $F$ is the number of factors, $k$ is the maximum degree of a factor, $l$ is the maximum degree of a variable, and $m$ is the maximum number of values a variable can take.

Now, we interpret (10) for the CGBG setting. In this case, $m = |\mathcal{A}_*|$, $F = O(\rho \cdot |\Theta_*|^k)$ is the number of contributions, and $l = O(\rho_* \cdot |\Theta_*|^{k-1})$, since each edge $e \in \mathcal{E}$ in which a (type of an) agent participates induces $O(|\Theta_*|^{k-1})$ contributions to which it is connected. □

The implication of this theorem it that a MAX-SUM iteration is tractable for small $k$, the size of the largest local scope. For problems with bounded $k$, this therefore directly yields a polynomial-time approximate algorithm by bounding the number of iterations. Given these results, we expect that, in general, MAX-SUM will prove more effective than NDP at exploiting type independence. In particular, we hypothesize that MAX-SUM will outperform NDP when applied to an ATI factor graph and will outperform MAX-SUM when applied to an AI factor graph.

---

[5] However, some variants of the message passing approach have slight modifications that yield convergence guarantees (Globerson and Jaakkola, 2008). Since we found that regular MAX-SUM performs well in our experimental setting, we do not consider such variants here.

## 5 Experiments

To evaluate the efficacy of using ATI factor graphs, we compare **NDP-ATI**, NDP applied to an ATI factor graph and **Max-Sum-ATI**, MAX-SUM applied to an ATI factor graph with 10 restarts, to several state-of-the-art alternatives. In particular, we compare to: **BaGaBaB**, Bayesian Game Branch and Bound, a heuristic search method for optimally solving CBGs (Oliehoek et al., 2010); **AltMax**, Alternating maximization, starting with a random joint policy, each agent iteratively computes a best-response policy for each of its types, thus hill-climbing to a local optimum; **CE**, Cross Entropy optimization (de Boer et al., 2005) a randomized optimization method that maintains a distribution over joint policies; **MAID-CM**, the state-of-the-art continuation method for solving multiagent influence diagrams (Blum et al., 2006).[6] Apart from MAID-CM, methods were implemented using the MADP Toolbox (Spaan and Oliehoek, 2008); the NDP and MAX-SUM implementations also use LIB-DAI (Mooij, 2008).

### 5.1 Random CGBG Experiments

We first present experiments in randomly generated games following a procedure similar to that used by Kok and Vlassis (2006). We start with a set of $n$ agents with no local payoff functions defined. As long as a path does not exist between every pair of agents, we add a local payoff function involving $k$ agents. Payoffs $u^e(\boldsymbol{\theta}_e, \mathbf{a}_e)$ are drawn from a normal distribution $\mathcal{N}(0, 1)$, and the local joint type probabilities $\Pr(\boldsymbol{\theta}_e)$ are drawn from a uniform distribution and then normalized.

These random BGs enable testing on a range of problem parameters. In particular, we investigate the effect of scaling the number of agents $n$, the number of types for each agent $|\Theta_i|$, and the number of actions for each agent $|\mathcal{A}_i|$. We assume that the scopes of the local payoff functions, the individual action sets, and the individual type sets all have the same size for each agent. For each set of parameters, we generate 1,000 random CGBGs. Each method is run on these instances, limited by 1GB of memory and $5s$ computation time.[7] Payoffs are normalized with respect to those of MAX-SUM-ATI (whose payoff is always 1).

First, we compare NDP-ATI and MAX-SUM-ATI with other methods that do not exploit a factor graph rep-resentation explicitly, the results of which are shown in Fig. 5. These graphs show the payoff and computation time when increasing the number of agents (Fig. 5a and 5d), the number of actions (Fig. 5b and 5e), and the number of types (Fig. 5c and 5f). These results support a number of observations. 1) As predicted, NDP scales well with respect to the number of agents, but not types. It also scales poorly with the number of actions, which indicates that the encountered induced width is quite high in practice. 2) Though NDP cannot successfully exploit type independence, it is still consistently faster than BAGABAB's heuristic search. 3) The approximate MAX-SUM method always finds the optimum when we can compute it. 4) MAX-SUM finds substantially better solutions than the other approximate methods ALTMAX and CE, while running faster than them for all but the smallest problems. 5) MAID-CM, the state-of-the-art for MAIDs, is significantly slower than the other methods.

---

[6] Since we use the implementation of Blum et al., which does not output the quality of the found solution, we report only computation times.

[7] A data point is not presented if the method exceeded the predefined resource limits on one or more test runs. MAID-CM was run on 30 games and given 30s.

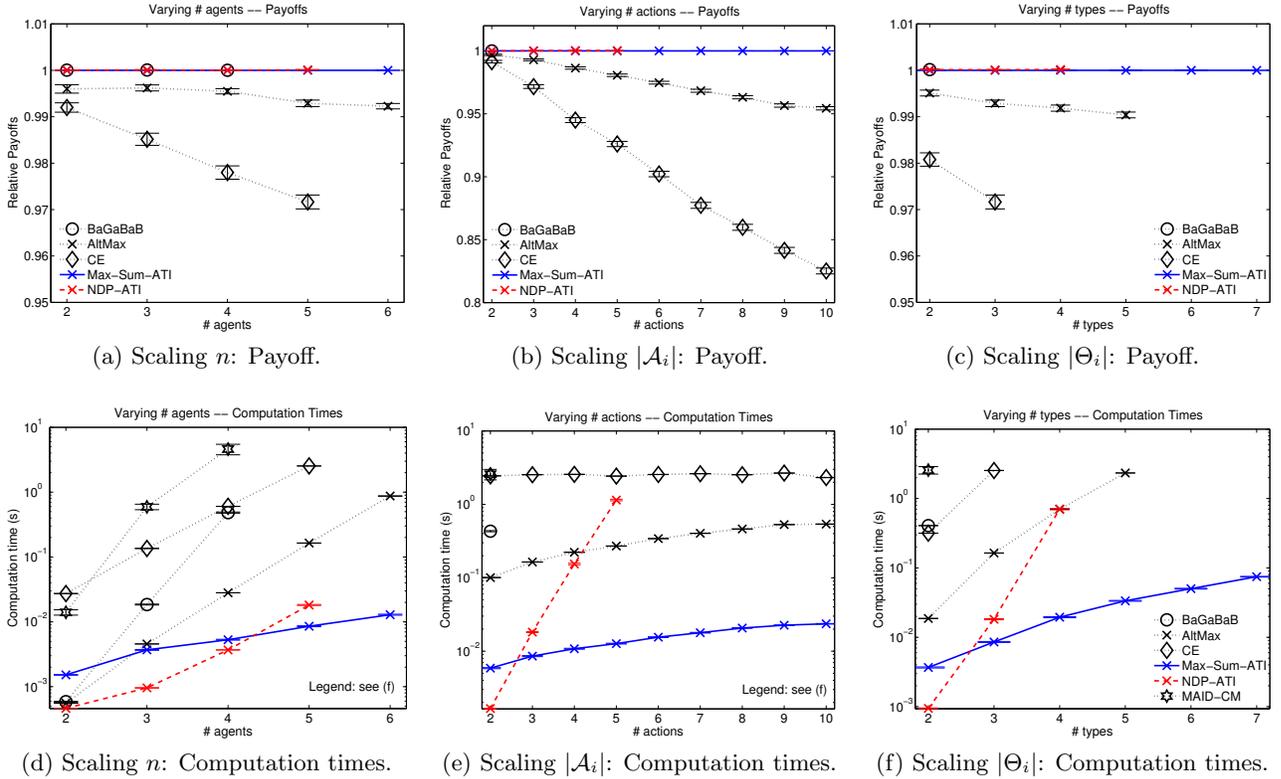

Figure 5: Comparison of MAX-SUM-ATI and NDP-ATI with other methods, scaling the number of agents ((a) and (d)), the number of actions ((b) and (e))), and the number of types ((c) and (f)). In all plots $k = 2$, and the parameters that are not being varied are set as follows: $n = 5, |\mathcal{A}_i| = 3, |\Theta_i| = 3$.

To make explicit the advantage of exploiting type independence, Fig. 6 compares the performance of MAX-SUM-ATI to that of MAX-SUM-AI, i.e., MAX-SUM applied to an AI factor graph, for games with $k = 2, |\Theta_i| = 4, |a_i| = 4$, larger numbers of agents, and $30s$ allowed per CGBG. MAX-SUM-AI scales only to 50 agents, while MAX-SUM-ATI scales to 725 (limited only by the allocated memory). While it is not possible to compute optimal solutions for these problems, values found by both methods are close and increase in proportion with the number of payoff components, indicating no degradation in performance.

While these results are for $k = 2$, experiments conducted for $k = 3$ were qualitatively similar, though lack of space prevents a detailed presentation. Note that there is no hope of efficiently solving problems with large $k$, since such problems cannot even be represented compactly. However, this not a serious practical limitation; on the contrary, even low values of $k$ allow for complicated interactions since each agent may participate in many local payoff functions.

### 5.2 Generalized Firefighting Experiments

In this section, we aim to demonstrate that the advantages of MAX-SUM-ATI extend to a more realistic problem, called GENERALIZED FIREFIGHTING, which is like the two-agent firefighting problem of Fig. 1 but with $N_H$ houses and $n$ agents physically spread over a 2-dimensional area. Each agent observes the $N_O$ nearest houses and may choose to fight fire at any of the $N_A$ nearest houses.

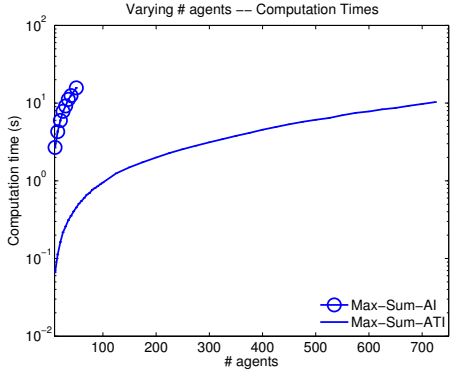

Figure 6: Comparison of the scaling behavior for many agents for MAX-SUM-AI and MAX-SUM-ATI.

Fig. 7 shows the results of our experiments in this domain, which corroborate our findings on random CGBGs by showing that MAX-SUM has the most desirable scaling behavior in a variety of different parameters.[8] We omit payoff plots because all methods computed solutions with the same payoff on all problem instances; MAX-SUM achieves the optimum in cases where we can compute it. Further investigation of solution quality (omitted due to lack of space) showed that solution quality decreases (there are only penalties in this problem) proportionally with the number of payoff components, again indicating that scalability does not come at the expense of solution quality.

## 6 Related Work

There is a large body of research that is related to the work presented in this paper. MAX-SUM has been used in decision making for teams of cooperative agents (Kok and Vlassis, 2006; Farinelli et al., 2008; Kuyer et al., 2008; Rogers et al., 2011). However, all of this work deals with cooperative games of perfect information, whereas we deal with imperfect information.

The CGBG model was introduced by Oliehoek et al. (2008b), who also used NDP to optimally solve CGBGs. However, NDP was applied only to the AI factor graph. Here we applied and analyzed application of NDP and MAX-SUM to a newly proposed ATI factor graph.

Two methods that are closely related to our approach are the heuristic search methods introduced by Oliehoek et al. (2010) and Kumar and Zilberstein (2010). The former propose BAGABAB for cooperative BGs, which we evaluated in our comparative experiments. It exploits additivity of the value function and can thus be interpreted as exploiting type independence. However, its performance depends heavily on the tightness of the heuristic used within BAGABAB (and, when problem sizes increase, the heuristic bounds tend to become looser, leading to disproportionate increases in runtime). Kumar and Zilberstein (2010) take a similar approach by performing search with a state-of-the-art heuristic (EDAC) for weighted constraint satisfaction problems to perform a point-based backup (as part of a solution method for sequential decision problems formalized as decentralized partially observable Markov decision processes, Dec-POMDPs). However, the EDAC heuristic provides leverage only in the two-agent case (de Givry et al., 2005). Furthermore, unlike our approach, neither of these methods exploits agent independence.

A closely related framework is the multiagent influence diagram (MAID), which extends decision diagrams to multiagent settings (Koller and Milch, 2003). A MAID represents a decision problem with a Bayesian network containing a set of chance nodes and, for each agent, a set of decision and utility nodes. As in a CGBG, the individual payoff function for each agent is defined as the sum of local payoffs. MAIDs are more general because they can represent sequential and non-identical payoff settings. Therefore, MAID solution methods are in principle applicable to CGBGs. However, these methods aim merely to find a sample Nash equilibrium, which is only guaranteed to be a local optimum. In contrast, our factor-graph formulation enables methods that find the global optimum. The MAID-CM method (Blum et al., 2006), which we also compared against, exploits the structure of the network in an inner loop to more efficiently compute the gradient of the payoff function. However, Blum et al. do not decompose the payoff into contributions and thus do not represent type independence. Moreover, since this inference is performed exactly using a clique-tree based approach, it cannot exploit type independence effectively, as the graphical argument made against NDP extends directly to this setting.

The framework of Bayesian action-graph games (BAGGs) can also model any BG (Jiang and Leyton-Brown, 2010). The BAGG solution method is similar to MAID-CM in that it exploits structure in an inner loop while searching for a sample equilibrium and thus suffers from the same limitations. This framework additionally allows the exploitation of anonymity and context-specific independence, but offers no advantages when these are not present.

---

[8] We omit NDP-ATI, since MAX-SUM-ATI outperformed it in random games, and MAID-CM, since it could not find solutions for any of these problems in 30$s$.

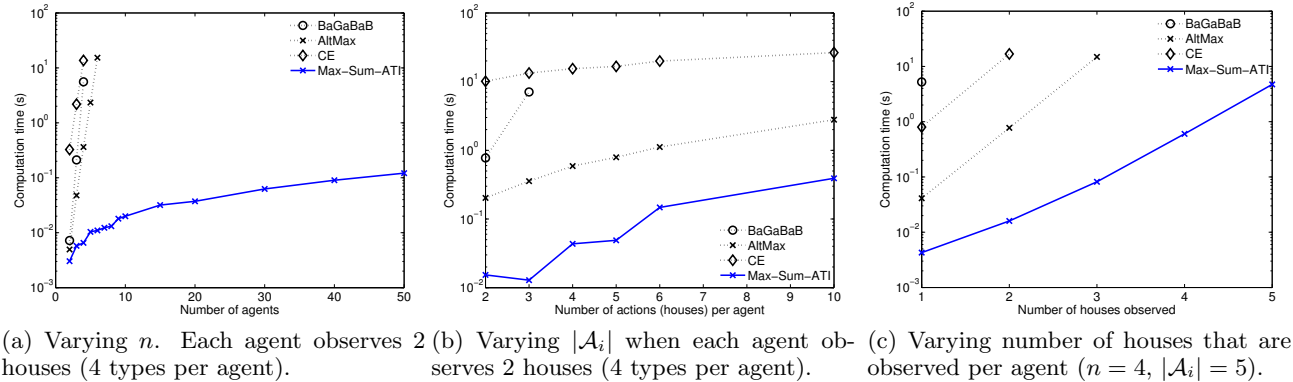

(a) Varying $n$. Each agent observes 2 houses (4 types per agent).
(b) Varying $|\mathcal{A}_i|$ when each agent observes 2 houses (4 types per agent).
(c) Varying number of houses that are observed per agent ($n=4$, $|\mathcal{A}_i|=5$).

Figure 7: Computation time results for the GENERALIZED FIRE FIGHTING problem.

## 7 Conclusions & Future Work

This paper considered the interaction of a team of agents under uncertainty modeled as cooperative graphical Bayesian games (CGBGs). We showed that CGBGs possess two different types of structure, agent and type independence, which can be neatly captured in a novel ATI factor graph formulation.

We considered two standard solution methods: non-serial dynamic programming (NDP) and MAX-SUM message passing. The former has a computational complexity that is exponential in the induced tree width of the factor graph, which we proved to be lower bounded by the number of individual types. One iteration of the latter is tractable when there is enough independence between agents: we showed that it is exponential only in $k$, the maximum number of agents that participate in the same local payoff function.

By bounding the number of iterations, the combination of the ATI factor graph and MAX-SUM yields, for any fixed $k$, a polynomial-time algorithm. While there are no quality guarantees, the complexity results suggest that any algorithm that scales polynomially with respect to the number of types (and thus, via our theorem, tree-width) *cannot* provide a solution within an arbitrary error bound (Kwisthout, 2011). It may be, however, be possible to obtain (non-adjustable) quality bounds for MAX-SUM solutions via a technique that eliminates edges from the factor graph (Rogers et al., 2011).

An empirical evaluation showed that solution quality is high: in all our experiments MAX-SUM found the optimal solution (when it could be computed). Moreover, the evaluation also revealed that exploiting both agent and type agent independence leads to significantly better performance than current state-of-the-art methods, providing scalability to much larger problems than was previously possible. In particular, we showed that this approach allows for the solution of coordination problems with imperfect information for up to 750 agents, limited only by a 1GB memory constraint. As such, we anticipate that the representation and exploitation of both agent and type independence will be a critical component in future solution methods for cooperative Bayesian games.

For future work, a possible extension of our approach could replace MAX-SUM with message passing algorithms for belief propagation that are guaranteed to converge (Globerson and Jaakkola, 2008). In certain cases, such approaches can be shown to compute an optimal solution (Jebara, 2009). A different direction of study would be to use our approach to help solve Dec-POMDPs, by replacing the BG solving components used by Oliehoek et al. (2008b) and Kumar and Zilberstein (2010) with NDP or MAX-SUM applied to the ATI factor graph. This can lead to huge scale-up in the number of agents approximate solutions can handle (Oliehoek, 2010). Similarly, it may be possible to improve methods like MAID-CM by performing the inference with MAX-SUM in order to exploit type independence. Another avenue is to extend our algorithms to the non-cooperative case by rephrasing the task of finding a sample Nash equilibrium as one of minimizing regret, as suggested by Vickrey and Koller (2002).

## Acknowledgements

We would like to thank Nikos Vlassis for extensive discussions, and Kevin Leyton-Brown, Mathijs de Weerdt, Christian Shelton, David Silver and Leslie Kaelbling for their valuable input. Research supported by AFOSR MURI project #FA9550-09-1-0538 and by NWO CATCH project #640.005.003. M.S. is funded by the FP7 Marie Curie Actions Individual Fellowship #275217 (FP7-PEOPLE-2010-IEF).